\documentclass[twocolumn,showpacs,showkeys,superscriptaddress]{revtex4}
\usepackage{amsmath}
\usepackage{amsfonts}
\usepackage{amssymb}
\usepackage{pbsi}
\usepackage[T1]{fontenc}
\usepackage{hyperref}
\usepackage{xcolor}

\def\openone{\leavevmode\hbox{\small1\kern-3.8pt\normalsize1}}
\def\N{\leavevmode\hbox{ Z \kern-8 pt\normalsize{Z}}}
\def\openone{\leavevmode\hbox{\small1\kern-3.8pt\normalsize1}}
\def\openJ{\leavevmode\hbox{J \kern-9.5pt\normalsize J}}
\def\openS{\leavevmode\hbox{ S \kern-9.3pt\normalsize S}}
\newcommand{\bb}{\begin{equation}}
\newcommand{\ee}{\end{equation}}
\newcommand{\eqb}{\begin{eqnarray}}
\newcommand{\eqf}{\end{eqnarray}}

\usepackage{color}

\begin{document}

\title{Quantum particles that behave as free classical particles}

\author{Sergio A. Hojman}
\email{sergio.hojman@uai.cl}
\affiliation{Departamento de Ciencias, Facultad de Artes Liberales,
Universidad Adolfo Ib\'a\~nez, Santiago 7491169, Chile.}
\affiliation{Centro de Investigaci\'on en Matem\'aticas, A.C., Unidad M\'erida, Yuc. 97302, M\'exico}
\affiliation{Departamento de F\'{\i}sica, Facultad de Ciencias, Universidad de Chile,
Santiago 7800003, Chile.}
\affiliation{Centro de Recursos Educativos Avanzados,
CREA, Santiago 7500018, Chile.}
\author{Felipe A. Asenjo}
\email{felipe.asenjo@uai.cl}
\affiliation{Facultad de Ingenier\'ia y Ciencias,
Universidad Adolfo Ib\'a\~nez, Santiago 7491169, Chile.}

\begin{abstract}
The existence of non--vanishing Bohm potentials, in the Madelung--Bohm version of the Schr\"odinger equation, allows for the construction of particular  solutions for states of quantum particles interacting with non--trivial external potentials that propagate as free classical particles. Such solutions are constructed with phases which satisfy the classical Hamilton--Jacobi for free particles and whose probability densities propagate with constant velocity, as free classical particles do.

\end{abstract}


\maketitle
 
 \section{Introduction}
 
In 1979, Berry and Balazs \cite{berry} showed that a quantum free wavepacket can show unexpected accelerating characteristics.
In this work we address the opposite question. That is,
whether the wavefunction of interacting particle, that satisfies the Schr\"odinger equation for a potential $V$, may propagate as free particles in the sense that their probability densities propagate with constant velocity as if they were free classical particles. In this work we establish the conditions which make this behavior possible and present numerous examples. 
The  theoretical results presented in this work are in the same spirit that several similar others that have had experimental confirmation in optics  and quantum mechanics \cite{mex,impens,sivichis,sivichis2,rivka,bloch}.

We prove that such possibility indeed exists in the framework of non--relativistic quantum mechanics and its related to the existence of the so--called Bohm potential. In other words, there are quantum solutions, for families of external potentials $V$, in which the wave function for the particle propagates as a free classical particle. This is only possible for a non--vanishing Bohm potential, which in turn implies that amplitude of the wavefunction is not  constant.
We focus in one--dimensional systems, although our results can be generalized to higher dimensions \cite{asenjoHojmanclasquandsper,sahfaz20201,makowski}, or to relativistic regime \cite{fanchi} following the ideas presented here. 

By free classical particle, we understand any particle of mass $m$ satisfying the free Hamilton--Jacobi (HJ) equation 
\begin{eqnarray}
\frac{1}{2m} \left(S'\right)^2 +\dot S&=&0\, . \label{HJBfree}
\end{eqnarray}
for an action $S=S(x,t)$, where  
$'\equiv\partial_x$, and $\dot{}\equiv\partial_t$. The propagating solutions of this equation have constant velocity.
Thus, this action may be considered as the phase of a solution to the Schr\"odinger equation. Therefore, we are looking for wavefunctions with a phase satisfying Eq.~\eqref{HJBfree}, and with an amplitude  that allow us to solve  Schr\"odinger equation for a given potential $V$.

Let us consider the wavefunction $\psi=\psi({x},t)$ of a one--dimensional Schr\"odinger equation  (and its complex conjugate) for a real potential $V({x},t)$
\begin{equation}
-\frac{{\hbar}^2}{2m}\psi'' + V \psi - i \hbar \dot\psi = 0\, .\label{schr}
\end{equation}
The wavefunction may be written in terms of a polar decomposition as ${{\psi}} = A\exp\left(i S/\hbar\right)$,
where the amplitude $A({ x},t)$ and the phase $S({ x},t)$ are real functions. Thereby, the Schr\"odinger equations become as \cite{madel,holland,wyatt,bohm,taka,durr}
\begin{eqnarray}
\frac{1}{2m} \left(S'\right)^2 +V_B + V +\dot S&=&0\, , \label{HJB1} \\  \frac{1}{m}  \left(A^2 \, S'\right)'  + \left ({A^2}
 \right){\dot \ }&=&0\, . \label{cont}    
\end{eqnarray}
where the Bohm 
potential is given by
\begin{equation}
V_B \equiv -\frac{\hbar^2}{2m}\frac{ A'' }{A}\ . \label{VB}    
\end{equation}
The first equation \eqref{HJB1} is the quantum Hamilton--Jacobi (QHJ) equation for the (external) potential $V$. The quantum modification consists in the addition of the Bohm potential to the classical HJ equation. The second equation \eqref{cont} is the continuity (probability conservation) equation. 
 To enforce that the probability density of a quantum interacting particle propagates as a free classical particle, we need to require that the Bohm potential cancels out any contribution of the external potential,
\begin{equation}
V_B +V=0\, .\label{VB2}    
\end{equation}
allowing the phase, from Eq.~\eqref{HJB1}, to fulfill the HJ equation \eqref{HJBfree}.
Above condition implies that the external potential determines completely the dynamics of the amplitude $A$, through the Bohm potential. This also must be consistent with the continuity equation \eqref{cont}.

The continuity equation \eqref{cont} is identically solved by defining the arbitrary potential function ${f}={ f}({ x},t)$, such that $A^2=f'$, and $A^2 S'=- m \dot{ f}$.
 For a one--dimensional system, once the free particle action $S$ is found by solving HJ equation \eqref{HJBfree}, ${f}={f}({x},t)$ can be determined by the relation
\begin{eqnarray}
f' S'+m \dot f=0\, .
\label{condicionparaf}
\end{eqnarray}
This equation states that $f$ depends on $x$ and $t$ through one variable only. On the other hand, the amplitude of the wavefunction is found to be given by the relation (with $f'>0$)
\begin{eqnarray}
A^2={f'}\, .
\label{amplitudedefinition}
\end{eqnarray}
This allows, in principle, to have negative amplitudes. Examples of this are shown below, as well as others where the amplitude is always positive (without nodes in $x$) and  they are also normalizable.

When the exact form of the amplitude $A$ (or function $f$) is found by solving Eq.~\eqref{VB2},  a quantum particle in the presence of a potential $V$ propagates as a free particle, in the sense that its phase is equal to the action for a free classical particle, while its amplitude have a $x$ and $t$ dependence completely determined by the phase. Therefore, this solution is a quantum state with a phase that coincides with the action for a free classical particle, while its probability density  propagates as a free classical particle with constant velocity. These solutions 
are nondiffracting
wavepackets and they do not correspond to a solution in a classical limit \cite{durr}, neither can be found by a Galilean transformation \cite{cervero}.

\section{Separable action for free classical particle}

Let us study the simplest case for a classical free particle, in which the spatial and temporal dependence are separated. The  phase (action) is given by
\begin{eqnarray}
S(x,t)= k\, x-\frac{k^2}{2m} t \, .
\label{freeplanewave}
\end{eqnarray}
This action is a solution of \eqref{HJBfree} for any constant $k$.
Eq.~\eqref{condicionparaf} allows us to find that $f$ depends on $x$ and $t$ through one variable $z$ only. In this case, we obtain that it has the form $f(x,t)={ f}\left(z\right)={f}\left(x-k t/m\right)$,
and thus, by Eq~\eqref{amplitudedefinition}, we obtain that the amplitude depends on the same variable $z$ as
\begin{equation}
    A(x,t)^2={ A}\left(z\right)^2=\frac{d f}{dz}\, ,
\end{equation}
where $z$ is defined as
\begin{equation}
    z\equiv x-\frac{k}{m} t\, .
\end{equation}

A quantum particle interacting with an external potential $V(x,t)=V(z)$, propagates  with phase \eqref{freeplanewave}, if the amplitude fulfills Eq.~\eqref{VB2}, in the form
\begin{eqnarray}
V\left(z\right)=\frac{\hbar^2}{2m { A}\left( z\right)}\frac{d^2{ A}\left(z\right)}{d z^2}\, .
\label{VB3}
\end{eqnarray}
For this case, all considered external potentials $V$ must depend on the $z$ variable, and therefore, they are not static. In this form, any solution of Eq.~\eqref{VB3} corresponds to a quantum particle which propagates with the action of a free classical particle, and its probability density moves with the velocity $k/m$.
Several different solutions are described below.

{\it Constant force.}
Consider a constant force $F=-V'$, with potential $V(z)= - F z$. Thus, Eq.~\eqref{VB3} produces an amplitude  given in terms of Airy functions 
\begin{eqnarray}
A(x,t)={\mbox{Ai}}\left(-\left(\frac{2m F}{\hbar^{2}}\right)^{1/3}z\right)\, .
\end{eqnarray}
This Airy wavepacket propagates as a free classical particle (without acceleration) under a constant force. 

{\it Moving potential trap.} An attractive potential with the form $V(z)=-\gamma\, \delta(z)$ is used to manipulate particles \cite{granot,mark}. Here $\gamma$ is a  constant, and $\delta$ is the Dirac delta function.
The amplitude solution of Eq.~\eqref{VB3} becomes
\begin{eqnarray}
A(x,t)=\frac{m\gamma\beta}{\hbar^2} z\,  {\mbox{sgn}}(z)-\beta \, ,
\end{eqnarray}
for an arbitrary constant $\beta$, and where ${\mbox{sgn}}$
is the sign function.

{\it Coulomb potential for a moving charge.} Let us assume a potential with the form $V(z)=\alpha/z$, for a moving charge with constant non--relativistic velocity ($\alpha$ is a constant). This corresponds to the non-relativistic expression for the Li\'enard--Wiechert four--potential  \cite{kurt}. In this case, Eq.~\eqref{VB3} gives an amplitude in terms of Bessel functions $K_1$
\begin{eqnarray}
A(x,t)=\frac{\sqrt{2m\alpha\, z}}{\hbar}\,  K_1\left(\frac{2\sqrt{2m\alpha\, z}}{\hbar} \right)\, .
\end{eqnarray}
Solutions in terms of Bessel functions $I_1$ are also possible. Thus, this Coulomb potential produces Bessel wavepackets that allows the particles propagate freely.

{\it Electromagnetic wave.} 
A particle  interacting with an electromagnetic wave (with wavenumber $\kappa$ and frequency $\kappa k /m$) experience a potential of the form $V(z)=\gamma \cos(\kappa z)$ (with constant $\gamma$). In this case, Eq.~\eqref{VB3} becomes a Mathieu equation
\begin{eqnarray}
\frac{d^2{ A} }{d z^2}-\frac{2m\gamma}{\hbar^2}\cos\left(\kappa z\right) A=0\, .
\end{eqnarray}
Explicit solutions are written in terms of the recurrence relations \cite{ruby,stegun}. In this form, Mathieu beam wavepackets support quantum solutions that propagate in a free classical fashion.

{\it Harmonic oscillator.} For a shifted harmonic oscillator $V(z)=m \omega^2 z^2/2$ \cite{duru}, with frequency $\omega$, Eq.~\eqref{VB3} has a solution Parabolic cylinder functions \cite{stegun}
\begin{eqnarray}
A(x,t)=D_{-\frac{1}{2}}\left(\sqrt{\frac{2 m\omega}{\hbar}} z\right)\, .
\end{eqnarray}

{\it P\"oschl--Teller potential.} Consider the  moving potential $V(z)=-\gamma\,  {\mbox{sech}}^2 z$, with constant $\gamma$.
 The amplitude solution of \eqref{VB3} is written in terms of a Legendre polynomial $P$ and a Legendre function of the second kind $Q$ as
 \begin{equation}
     A(x,t)=a_1 P_n\left(\tanh z\right)+a_2 Q_n\left(\tanh z\right)
 \end{equation}
 with arbitrary $a_1$ and $a_2$, and $n=(\sqrt{1+8m\gamma/\hbar^2}-1)/2$.

{\it Constant modified harmonic oscillator.} The above examples have the feature that the wavefunctions  have nodes
in $x$. However, normalizable states without nodes can also be found satisfying Eq.~\eqref{VB3}. An example of this is the system subject to a (shifted) harmonic oscillator, with  $V(z)=m \omega^2 z^2/2-\hbar \omega/2$, and arbitrary frequency $\omega$. This potential solves Eq.~\eqref{VB3} for the amplitude
\begin{equation}
    A(x,t)=\left(\frac{m\omega}{\hbar\pi}\right)^{1/4}\exp\left(-\frac{m\omega}{2\hbar} z^2\right)\, ,
\end{equation}
 which gives rise to a square--integrable  wavefunction.

{\it Constant modified P\"oschl--Teller potential.}
Examples of physical states
can also be obtained for the following 
P\"oschl--Teller potential $V(z)=-(\hbar^2/m)\, {\mbox{sech}}^2 z+\hbar^2/(2m)$.
In this case, solution of Eq.~\eqref{VB3} gives a propagating solitonic amplitude
\begin{equation}
    A(x,t)=\frac{1}{\sqrt{2}}\, {\mbox{sech}}\, z\, ,
\end{equation}
which produces a square--integrable
wavefunction with probability amplitude that propagates as a free classical particle with action \eqref{freeplanewave}.

\section{Non--separable action for free classical particle}

Another very well--known solution for the classical HJ \eqref{HJBfree} for classical free particles is
\begin{equation}
    S(x,t)=\frac{m\, (x-x_0)^2}{2(t-t_0)}\, ,
    \label{action2free}
\end{equation}
for arbitrary initial position $x_0$ and initial time $t_0<t$.
This action is a non--separable function of space and time.

In this case, Eq.~\eqref{condicionparaf} allows us to find that any function with the functionality $f(x,t)=f(y)$ solves the continuity equation,  where we have introduced  the variable
\begin{eqnarray}
y\equiv\frac{x-x_0}{t-t_0}\, .
\end{eqnarray}
Therefore, amplitude is given by
\begin{equation}
    A(x,t)=\frac{1}{\sqrt{t-t_0}}{\cal A}(y)\, ,
    \label{amplinonsperable}
\end{equation}
with ${\cal A}^2= {d f/dy}$.
In this case, any external potential with the form
\begin{eqnarray}
V(x,t)=\frac{1}{\left(t-t_0\right)^2}{\cal V}(y)\, ,
\label{generalforlpotentdifu}
\end{eqnarray}
allow us to re--write Eq.~\eqref{VB2} as
\begin{equation}
    {\cal V}(y)=\frac{\hbar^2}{2m {\cal A}(y)} \frac{ d^2 {{\cal A}(y)}}{dy^2}\, .
    \label{VB4}
\end{equation}

Potentials with the exact space and time dependence of the form
\eqref{generalforlpotentdifu}, have been shown to produce exact Feynman propagators \cite{dodonov}.  In this form, any solution of Eq.~\eqref{VB4} produces quantum particle that propagates classically with action \eqref{action2free} and probability density that propagates with constant velocity $(x-x_0)/(t-t_0)$ .
Below we study some of them in our context.

{\it Time decreasing force.} For a force decreasing in time with the form $F(t)=F_0/(t-t_0)^3$,  a potential ${\cal V}(y)=-F_0\, y$ can be used. In this case, Eq.~\eqref{VB4} produces Airy solutions, and amplitude \eqref{amplinonsperable} is
\begin{eqnarray}
A(x,t)=\frac{1}{\sqrt{t-t_0}}{\mbox{Ai}}\left(-\left(\frac{2m F_0}{\hbar^2}\right)^{1/3}y\right)\, .
\end{eqnarray}
Thus, for such forces, the quantum system is solved exactly, and the particle propagates as it were free.

{\it Harmonic oscillator.}  Consider the harmonic oscillator potential $V=m\omega^2 x^2/2$.
For a time--decreasing frequency in the form $\omega=\omega_0/(t-t_0)^2$ \cite{dodonov} (with constant $\omega_0$), then the harmonic oscillator with potential ${\cal V}(y)=m\omega_0^2 y^2/2$ can be solved exactly. Using Eq.~\eqref{VB4}, amplitudes are given in terms of Parabolic Cylinder functions \cite{stegun}
\begin{eqnarray}
A(x,t)=\frac{1}{\sqrt{t-t_0}} D_{-\frac{1}{2}}\left(\sqrt{\frac{2 m\omega_0}{\hbar}} y\right)\, .
\end{eqnarray}

{\it Coulomb--like potentials.} Consider a  potential with the form $V(x,t)=Z(t)/x$. When $Z$ decrease in time as $Z(t)=Z_0/(t-t_0)$ \cite{dodonov}, then ${\cal V}=Z_0/ y$, and there exist solutions using our approach. The amplitude of the wavefunction is again given in terms of Bessel functions $K_1$
\begin{eqnarray}
A(x,t)=\frac{\sqrt{2mZ_0\, y}}{\hbar \sqrt{t-t_0}}\,  K_1\left(\frac{2\sqrt{2mZ_0\, y}}{\hbar} \right)\, .
\end{eqnarray}

{\it Constant modified harmonic oscillator.}
As well as cases in Sec. II, 
square--integral wavefuntions can be obtaining satisfying Eq.~\eqref{VB4}. A simple example of this is for the harmonic oscillator with potential $V(x,t)=m \omega^2 x^2/2- \hbar\omega/2$,
with time--decreasing frequency  $\omega=\omega_0/(t-t_0)^2$. This allows us to define 
the potential ${\cal V}(y)=m \omega_0^2 y^2/2- \hbar\omega_0/2$, which according to Eq.~\eqref{VB4}, produces an amplitude of the form
\begin{equation}
    A(x,t)=\left(\frac{m\omega_0}{\hbar\, \pi\left(t-t_0\right)^2}\right)^{1/4}\exp\left(-\frac{m\omega_0}{2\hbar} y^2\right)\, .
\end{equation}
This amplitude defines a physical normalizable wavefunction that propagates with action \eqref{action2free} under this harmonic oscillator potential.

\section{Discussion}

With the above several examples and calculations we have shown that is possible for interacting quantum particles  to have a probability density that propagates as a free classical particle for a wide range of known potentials.
These quantum solutions have a phase that coincides with the action for a free classical particle, and therefore, they are non--trivial solutions of quantum mechanics.
This is only achieved because the Bohm potential of the wavefunction cancels out the external potential. By doing this, the external potential completely determines the amplitude of the wavepackets, as it can be seen in Eqs.~\eqref{VB3} and \eqref{VB4}.

The condition \eqref{VB2} allow us to describe our solutions as quantum particles that propagate as classical ones, as they satisfy the HJ equation \eqref{HJBfree}. However, they also satisfy the free-space Liouville equation $\partial_t {\mathcal F}+(p/m)\partial_x {\mathcal F}=0$, for the phase-space density \cite{phol}
\begin{equation}
{\mathcal F}(x,p,t)=A(x,t)^2\delta\left(p-\frac{\partial S}{\partial x}\right)\, ,
\end{equation}
where $\delta$ is the Dirac delta function, and $A^2=\int {\cal F} dp$ is the probability density of the studied solutions in each section. Here $p=\partial_x S$ is the constant momentum, which for solutions of Sec.~II is $p=k$, while for solutions of Sec.~III is $p=m(x-x_0)/(t-t_0)$. The solutions presented above  satisfy the free-space Liouville equation which implies that they behave as free classical particles.

It is remarkable the solutions explored along this work occur for the large family of potentials treated here. We think that they have remained largely unexplored and they can bring new insights in the propagation of quantum particles, as the quantum characteristics remain confined to the amplitude, while the phase is associated to the action of a free classical particle.
Furthermore, we have shown that square--integrable wavefunctions can be obtained for known potentials, thus representing physical states.

Any solution fulfilling condition \eqref{VB2} can now be interpreted as a nondiffracting wavepacket that modified its own probability density in order to propagate as if it were free. The implications of this behavior are not difficult to be envisaged as very interesting, as other quantum wavepackets, with similar features, such as accelerating and curved properties, have been constructed and measured in laboratories \cite{mex,impens,sivichis,sivichis2,rivka,bloch}.

\begin{acknowledgements}
  We thank an anonymous referee for all the comments that helped to improve this work.
\end{acknowledgements}



\begin{thebibliography}{17}

\bibitem{berry} M. V. Berry and N. L. Balazs, Am. J. Phys. {\bf 47}, 264 (1979).

\bibitem{mex} S. Ch\'avez-Cerda, U. Ruiz, V. Arriz\'on and H. M. Moya-Cessa, Opt. Exp. {\bf 19}, 16448 (2011).
\bibitem{impens} F. Impens, R. Duboscq, and D. Gu\'ery-Odelin, Phys. Rev. Lett. {\bf 124}, 250403 (2020).
\bibitem{sivichis} G. A. Siviloglou, J. Broky, A. Dogariu and D. N. Christodoulides, Phys. Rev. Lett. 99, 213901 (2007).
\bibitem{sivichis2} D. N. Christodoulides, N. K. Efremidis, P. Di Trapani and B. A. Malomed, Opt. Lett. {\bf 29}, 1446 (2004).
\bibitem{rivka} A. Patsyk, M. A. Bandres, R. Bekenstein, and M. Segev, Phys. Rev. X {\bf 8}, 011001 (2018).
\bibitem{bloch} N.  Voloch-Bloch, Y. Lereah, Y. Lilach, A. Gover and A. Arie, Nature {\bf 494}, 331 (2013).

\bibitem{asenjoHojmanclasquandsper} S. A. Hojman and F. A. Asenjo, Phys. Scr. {\bf 95} 085001 (2020).
\bibitem{sahfaz20201} S.A. Hojman and F.A. Asenjo, Phys. Lett. A {\bf 384}, 126263 (2020).
\bibitem{makowski} A. J. Makowski and S. Konkel,Phys. Rev. A {\bf 58}, 4975 (1998).
\bibitem{fanchi} J. R. Fanchi, Found. Phys. {\bf 30}, 1161 (2000).
\bibitem{madel} E. Madelung, Z. Physik {\bf 40}, 322 (1927).
\bibitem{bohm} D. Bohm, Phys. Rev. {\bf 85}, 166 (1952).
\bibitem{holland} P. R. Holland, {\it The Quantum Theory of Motion: an account of the de Broglie-Bohm causal interpretation of quantum mechanics}, (Cambridge University Press, 1993).
\bibitem{wyatt} R. E. Wyatt, {\it Quantum Dynamics with Trajectories: introduction to quantum hydrodynamics} (Springer, 2005).
\bibitem{taka} T. Takabayasi, Prog. Theor. Phys. {\bf 14}, 283 (1955); {\bf 12}, 810
(1954); {\bf 13}, 222 (1955); Phys. Rev. {\bf 102}, 297 (1956).
\bibitem{durr} D. D\"urr, S. Goldstein and N. Zanghi, {\it Quantum Physics without Quantum Philosophy} (Springer, 2013).
\bibitem{cervero} J. M. Cerver\'o and P. P. Polo, Eur. J. Phys. {\bf 37}, 055401 (2016).

\bibitem{granot} E. Granot and A. Marchewka, EPL {\bf 86}, 20007 (2009).
\bibitem{mark} M. R. A. Shegelski,T. Poole and C. Thompson, Eur. J. Phys. {\bf 34}, 569 (2013).
\bibitem{kurt} K. Lechner, {\it Classical Electrodynamics:A  Modern Perspective} (Springer, 2018).
\bibitem{ruby} L. Ruby, Am. J. Phys. {\bf 64}, 39 (1996).
\bibitem{stegun} M. Abramowitz and I. A. Stegun, {\it Handbook of Mathematical Functions with Formulas, Graphs, and Mathematical Tables}, (National Bureau of Standars, Applied Mathematics Series, Tenth Printing, 1972).
\bibitem{duru} I. H. Duru, J. Phys. A: Math. Gen. {\bf 22}, 4827 (1989).
\bibitem{dodonov}  V. V. Dodonov, V. I. Man'ko and D. E. Nikonov, Phys. Lett. A {\bf 162}, 359 (1992).
\bibitem{phol} P. Holland, Nuovo Cimento {\bf 116}, 1043 (2001).



\end{thebibliography}
\end{document}